\documentclass[showpacs,showkeys,
preprintnumbers,amsmath,amssymb,
superscriptaddress,
notitlepage,
]{revtex4-1}

\usepackage{amssymb}
\usepackage{amsmath}
\usepackage{graphicx}
\usepackage{color}
\usepackage{hyperref}
\usepackage{subfigure}
\usepackage{amssymb}
\usepackage{enumerate}
\usepackage{amsthm,amsfonts,amsmath}
\usepackage{bm}
\usepackage{graphicx}
\usepackage{graphics,epsf,psfrag,pifont,dsfont,bbold}
\usepackage{siunitx}
\usepackage{color}
\usepackage{xspace}
\usepackage{accents}
\usepackage[utf8]{inputenc}
\usepackage[T1]{fontenc}
\usepackage{soul}

\newcommand{\be}{\begin{equation}}
\newcommand{\ee}{\end{equation}}
\newcommand{\ben}{\begin{equation*}}
\newcommand{\een}{\end{equation*}}

\newcommand{\tshear}{\tau_{\mbox{\tiny {shear}}}}

\newcommand{\Cabar}{\overline {\mbox{Ca}}}
\newcommand{\Ca}{\mbox{Ca}}

\renewcommand{\i}{\textup{i}}

\newcommand{\delF}[1]{\textcolor{blue}{}}

\begin{document}



\title{Droplet breakup driven by shear thinning solutions in a microfluidic T-junction}


\author{Enrico Chiarello}
\affiliation{Dipartimento di Fisica e Astronomia ``Galileo Galilei'' - DFA, Universit\`a di Padova, Via F. Marzolo 8, 35131 Padova, Italy}

\author{Anupam Gupta}
\altaffiliation{Present address: Mechanical Science and Engineering, University of Illinois, 1206 W. Green Street, Urbana, IL 61801, USA}
\affiliation{Dipartimento di Fisica and INFN, Universit\`a di Roma 2 ``Tor Vergata'', Via della Ricerca Scientifica 1, 00133 Roma, Italy}

\author{Giampaolo Mistura}
\affiliation{Dipartimento di Fisica e Astronomia ``Galileo Galilei'' - DFA, Universit\`a di Padova, Via F. Marzolo 8, 35131 Padova, Italy}

\author{Mauro Sbragaglia}
\email[]{sbragaglia@roma2.infn.it}
\affiliation{Dipartimento di Fisica and INFN, Universit\`a di Roma 2 ``Tor Vergata'', Via della Ricerca Scientifica 1, 00133 Roma, Italy}

\author{Matteo Pierno}
\email[]{matteo.pierno@unipd.it}
\affiliation{Dipartimento di Fisica e Astronomia ``Galileo Galilei'' - DFA, Universit\`a di Padova, Via F. Marzolo 8, 35131 Padova, Italy}



\begin{abstract}Droplet-based microfluidics turned out to be an efficient and adjustable platform for digital analysis, encapsulation of cells, drug formulation, and polymerase chain reaction. Typically, for most biomedical applications, the handling of complex, non-Newtonian fluids is involved, e.g. synovial and salivary fluids, collagen, and gel scaffolds. In this study we investigate the problem of droplet formation occurring in a microfluidic T-shaped junction, when the continuous phase is made of shear thinning liquids.
At first, we review in detail the breakup process providing extensive, side-by-side comparisons between Newtonian and non-Newtonian liquids over unexplored ranges of flow conditions and viscous responses. The non-Newtonian liquid carrying the droplets is made of Xanthan solutions, a stiff rod-like polysaccharide displaying a marked shear thinning rheology. By defining an effective Capillary number, a simple yet effective methodology is used to account for the shear-dependent viscous response occurring at the breakup. The droplet size can be predicted over a wide range of flow conditions simply by knowing the rheology of the bulk continuous phase. Experimental results are complemented with numerical simulations of purely shear thinning fluids
 using Lattice Boltzmann models. The good agreement between the experimental and numerical data confirm the validity of the proposed rescaling with the effective Capillary number.
\end{abstract}


\maketitle

%
\section{Introduction}\label{sec:intro}
In the past decade, droplet-based microfluidics has been successfully applied to high-throughput chemical and biological analysis, synthesis of advanced materials, sample pretreatment, protein crystallization, encapsulation of cells and digital PCR systems~\cite{demello_control_06,teh08,baroud_dynamics_2010,theberg10,Seeman12,elvira_13,ferraro2016microfluidic,hindson2011high}. It is common to refer to the liquid forming the droplets as the dispersed phase, which is carried by the stream of a second, immiscible fluid identified as the continuous phase. Various approaches are adopted to produce uniform trains of droplets, including breakup in co-flowing streams, breakup in stretching- or elongational-dominated flows, and breakup in cross-flowing streams~\cite{christopher_microfluidic_2007,baroud_dynamics_2010}.\\
The process of droplet formation is relatively well understood and studied in the case of two immiscible Newtonian fluids, for example, water and oil~\cite{thorsen_dynamic_2001,Anna_apl_03,cramer04,Garstecki06,piccin2014generation}. More recently, these studies have been extended to non-Newtonian liquids because of an increasing interest in non-Newtonian multiphase microsystems~\cite{ren2015breakup}. These involve physiological fluids~\cite{yager2006microfluidic} such as blood (including fibrinogen for fibrin formation~\cite{pierno2006fbsa,retzinger1998adsorption}), synovial or salivary fluids~\cite{wagli2013microfluidic}, as well as fluid jets used in printing and spraying technology~\cite{bienia2016inkjet,mckinley2005visco}, and food emulsions~\cite{muijlwijk2016cross,FuTaoTao2015}. Most of the attention has been so far devoted to the formation of non-Newtonian droplets carried by Newtonian continuous phases~\cite{Steinhaus,christopher_passive_2009}. For elastic polymers, the effect of the molecular weight on filament thinning has been clarified in flow-focusing devices~\cite{Arratia08,Arratia09}. Only a few studies consider droplets carried by a non-Newtonian medium, using either flow focusing geometries~\cite{Garstecki13,ren2015breakup}, or  air bubbles formation~\cite{fuTaoTao2015bubble}.
\\
In the present study we address the generation of droplets by shear thinning continuous phases in T-junctions. We show how to describe this breakup in terms of an effective Capillary number, i.e. a standard way to measure the importance of interfacial forces with respect to (shear-dependent) viscous forces.
As non-Newtonian fluids, we investigate solutions of Xanthan, a stiff rod-like polysaccharide exhibiting a predominant shear thinning behavior and weak elastic effects~\cite{epjesliding,whitcomb78Xanth}.
The droplet size is analyzed as a function of the flow properties in various dynamical regimes. Because of the shear thinning effects, the viscosity is non homogeneous in space and varies with the flow rates of the two phases.
By quantitatively comparing Newtonian and non-Newtonian data, robust experimental evidence is provided that the droplet size rescales nicely with an effective Capillary number $(\Cabar)$, which reduces to the usual Capillary number $(\Ca)$ when both liquids are Newtonian. Experiments are complemented with numerical simulations of purely thinning fluids based on the lattice Boltzmann models (LBM), which are in good agreement with the experimental data and confirm the proposed scaling.

The paper is organized as follows: In Sec.~\ref{sec:matmed}, we describe the experimental (\ref{sec:experiments}) and numerical (\ref{sec:numerics}) methodologies, including the liquids, the T-shaped junctions and the definition of the effective Capillary number. Results are shown and discussed in Sec.~\ref{sec:experimental_results}, by reporting the size of the droplets (\ref{sec:size}), the rescaling of the size over the effective Capillary number (\ref{sec:rescaling}), and the distribution of both the velocity and viscous stress in the droplet-carrying phase along the T-junction (\ref{sec:numerical_results}). Conclusions and final remarks are found in Sec.~\ref{sec:conclusions}.

%
\section{Materials and Methods}\label{sec:matmed}
%
\begin{table*}[t!]
  \small  
   \begin{tabular*}{1.0\textwidth}{@{\extracolsep{\fill}} l l l l l l l}
    \hline
    \textbf{ID} & \textbf{Dispersed Phase} & \textbf{Newtonian Continuous Phase} & $\mathbf{\sigma (mN/m)}$ & $\mathbf{\eta_{d} (mPa\: s)}$ & $\mathbf{\eta_{c} (mPa\: s)}$  & $\mathbf{\lambda}$  \\
    \hline  
    
    N1 & Soybean Oil &  Water + 0.50\% Triton X-100 & $2.55\pm 0.03$ & $49.1$ & 0.9 & $\sim 50$ \\

    N2 & Soybean Oil & Glycerol/Water 60\% + 0.56\% Triton X-100 &  $3.59\pm 0.62$ & $49.1$ & 9 & $\sim 5$  \\

     N3 & Glycerol/Water 67\% & Hexadecane + 1\% Span 80& $3.99\pm 0.43$ & 14.5 & 3  & $\sim 5$  \\
     
     N4& Glycerol/Water 40\% & Hexadecane + 1\% Span 80 & $4.17\pm 0.09$  & 3.22 & 3  & $\sim 1$ \\
     
     N5 & Water & Hexadecane + 1\% Span 80& $5.01\pm 0.33$ & 0.9 & 3  & 0.3  \\
    \hline
   \end{tabular*}
   \caption{Experimental parameters of the liquids of the Newtonian systems. All quantities refer to a temperature $T=25^{\circ}C$. Surfactants and Glycerol concentration are expressed in terms of weight/weight ratio.} 
   \label{tbl:expNewton}
\end{table*}
\begin{table*}[t!]
  \small  
   \begin{tabular*}{1.0\textwidth}{@{\extracolsep{\fill}} l l l l l l l}
    \hline
       \textbf{ID} & \textbf{Dispersed Phase} & \textbf{shear thinning Continuous Phase} & $\mathbf{\sigma (mN/m)}$ & $\mathbf{\eta_{d} (mPa\: s)}$ & $\mathbf{\textit{K} (mPa\: s^{\textit{n}})}$ & $\mathbf{\textit{n}}$  \\
     
     \hline
     X400 & Soybean Oil & Xanthan 400~ppm + 0.2\% Triton X-100 &  $3.42\pm 0.01$ & $49.1$ & $32.6$ & $0.589$  \\
     
     X800 & Soybean Oil & Xanthan 800~ppm + 0.2\% Triton X-100 &  $3.00\pm 0.04$ & $49.1$ & $75.5$ & $0.491$  \\
     
     X1500 & Soybean Oil & Xanthan 1500~ppm + 0.7\% Triton X-100 &  $2.28\pm 0.02$ & $49.1$ & $312.5$ & $0.389$  \\
     
    \hline
   
   \end{tabular*}
   \caption{Experimental parameters of the liquids of the non-Newtonian systems. All quantities refer to a temperature $T=25^{\circ}C$. The concentrations are expressed in terms of weight/weight ratio.} \label{tbl:expThinning}
\end{table*}
%
%
%
\subsection{Experiments}\label{sec:experiments}
\begin{figure*}[htb]
    \begin{minipage}[]{0.49\textwidth}
\includegraphics[width = 0.85\columnwidth,trim=0 0 0 0,clip]{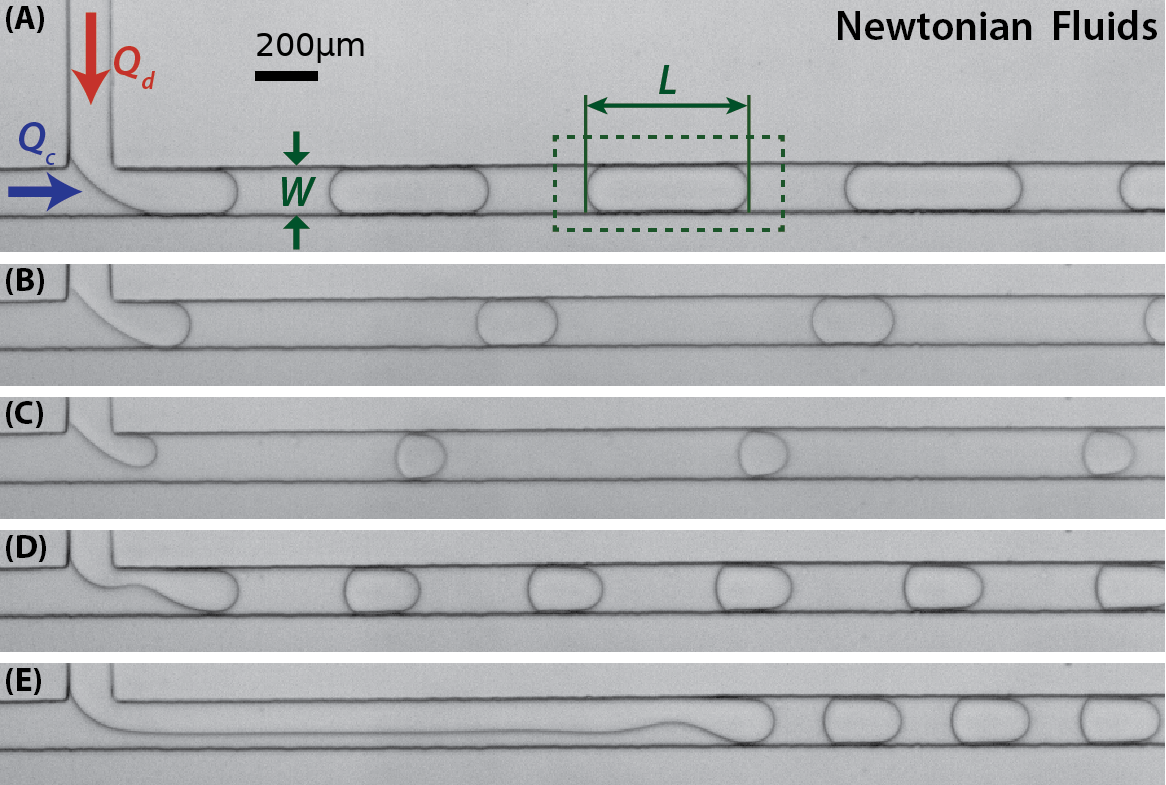}
\end{minipage}\hfill
\begin{minipage}[c]{0.49\textwidth}
\includegraphics[width = 0.85\columnwidth,trim=0 0 0 0,clip]{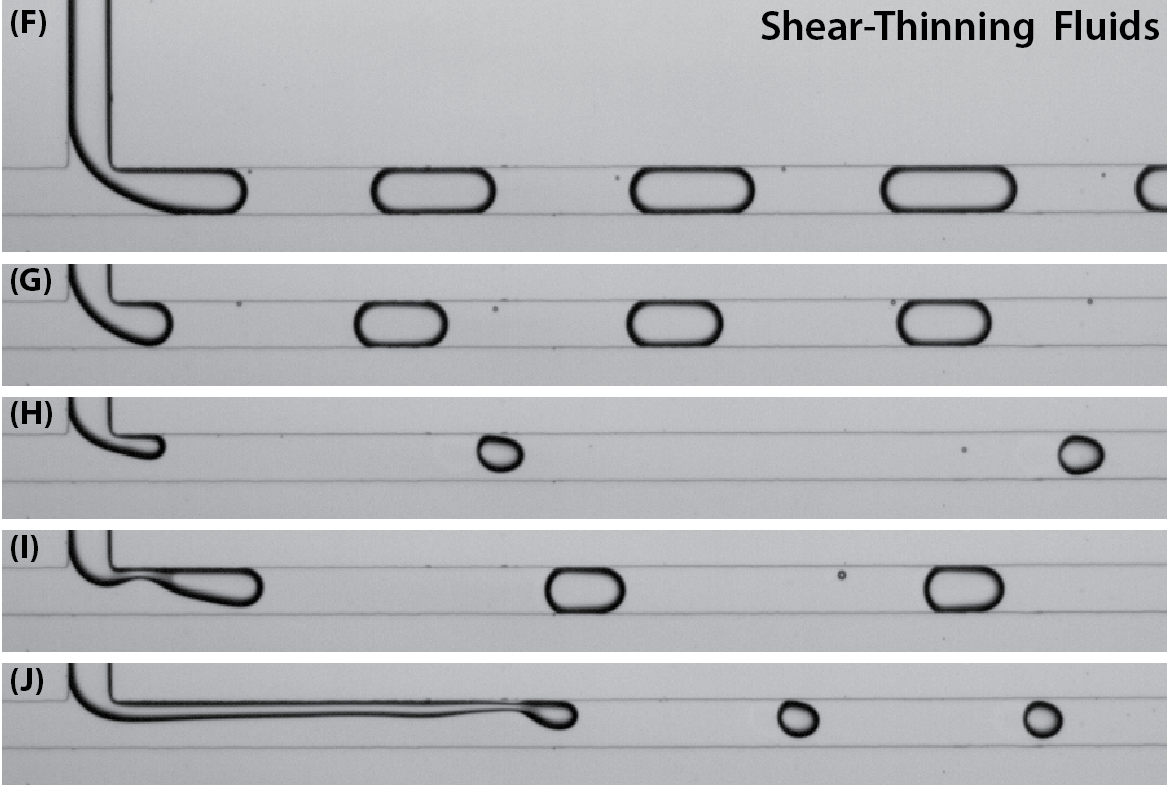}
\end{minipage}
\caption{Images of droplet breakup at the microfluidic T-junction in the Newtonian (left, liquid N3 of Table~\ref{tbl:expNewton} is shown) and shear thinning (right, X400 of Table~\ref{tbl:expThinning} is shown) continuous phase. The inlets of the junction have the same width $W$ and the same height;
the dispersed phase enters in the junction from the top with flow rate $Q_d$, the continuous phase comes from the left side with flow rate $Q_c$. The length $L$ of the droplets is measured by real time image-processing while they cross a rectangular window (the dashed contour) positioned downstream of the junction. 
Droplets are formed in squeezing (see snapshots A,F), dripping (C,H), or jetting (E,J) regimes. Snapshots (B,G) and (D,I) outline the emergent dripping and jetting regime respectively.
}
\label{fig:exp_snaps}
\end{figure*}

Droplets are produced by merging two immiscible liquids in a microfluidic
T-junction (Fig.~\ref{fig:exp_snaps}-A), which is composed of a main
microchannel encountering perpendicularly a side channel, having the same
cross section. 
The chips are made in polydimethylsiloxane (PDMS, Sylgard 184, Dow Corning) using standard photo-softlithography~\cite{ferraro2012morphological,toth2011suspension} and have an overall size of $2\, \times 5\,\mathrm{cm}$. The microchannels have a width $W \approx 150~\mu m$ and a height $H \approx 100~\mu m$.
Further details about the fabrication can be found in Sec.~I of the Supplemental Material (SM)~\cite{suppmat}.

The dispersed phase forming the droplets is injected by the side channel with a flow rate $Q_d$, while the continuous phase carrying away the
droplets is injected in the main channel with a flow rate $Q_c$, the
flow rates being controlled by a couple of syringe pumps specifically
designed for viscous flows (PHD~2000 from Harvard Apparatus, USA). 
Images of the droplets are acquired and analyzed in  with custom
 software, which also controls the syringe pumps. The length $L$ of the
droplets is measured, after breakup, in a region of interest downstream of
the T-junction (Fig.\ref{fig:exp_snaps}-A), and is averaged over at least a
hundred droplets~\cite{chiarello2015generation}.
Additional details on the experimental setup are reported in Sec.~II of the SM~\cite{suppmat}.
%
%
%
%
We address the generation of droplets both in Newtonian continuous phases (Newtonian systems) and
in shear thinning continuous phases made of polymer solutions (non-Newtonian systems).
Table~\ref{tbl:expNewton} summarizes the Newtonian fluids reporting the
liquids composition, interfacial tension $\sigma$, dynamic viscosity $\eta_d$ of the dispersed phase,
dynamic viscosity $\eta_c$ of the continuous phase and the viscosity ratio $\lambda=\eta_d/\eta_c.$ 
The experiments span two decades of $\lambda$ using various combinations of
either Soybean oil (Alfa Aesar) or Hexadecane (Sigma Aldrich), and water
solutions of Glycerol ($\geq$ 99.5\% anhydrous, Sigma Aldrich) at different
concentrations. A surfactant is added to the continuous phase in order to
improve the wetting of the channel walls. Triton~X-100 (Sigma Aldrich) is
used in the water solutions while Span~80 (Sigma Aldrich) is added to
Hexadecane. The values of the interfacial tension $\sigma$ are measured with
the pendant drop technique~\cite{chiarello2015generation}.
The non-Newtonian systems are summarized in Table~\ref{tbl:expThinning}.
Soybean oil is used as the dispersed phase while, as continuous phase, water
solutions of Xanthan (molecular weight $M_w \simeq 10^6$ g/mol, Sigma
Aldrich) at different concentrations are employed.
The rheology of Xanthan solutions used in the present study is discussed in Ref.~\cite{epjesliding}. Details can be found in Fig.~S2 of the SM~\cite{suppmat}. Briefly, at the concentrations considered, they exhibit a well pronounced shear thinning behavior and weak elastic effects due to the emergence of first normal stress differences at relatively high concentrations~\cite{epjesliding,whitcomb78Xanth}. The viscosity data of the Xanthan solutions are fitted according to the power law fluid model, similarly to that done in Refs.~\cite{epjesliding,chiarello2015generation}, 
\begin{equation}\label{eq:PowerLawFluidtext}
\eta (\dot{\gamma}) = K \, \dot{\gamma}^{\; (n-1)},
\end{equation}
where $K$ and $n$, being the fluid consistency and the flow behavior index respectively~\cite{macosko94}, are used as fitting parameters. Their values are listed in Table~\ref{tbl:expThinning} for the concentrations of Xanthan used in this study.
In the breakup process occurring in a microfluidic T-junction, the droplet size is usually analyzed in terms of the Capillary number~\cite{Demenech07,Gupta10,LiuZhang09b,Garstecki06,Seeman12}: 
\begin{equation}\label{eq:Catext}
\mbox{Ca}=\frac{\eta_c U_{av}}{\sigma},
\end{equation}
where the quantities $U_{av}$ and $\eta_c$ refer to the average velocity and the viscosity of the continuous phase, respectively, while $\sigma$ is the interfacial tension between the two phases. In the case of a non-Newtonian continuous phase, $\eta_c$ is chosen to be the value corresponding to an effective shear rate:
\begin{equation}\label{eq:shear_av}
\dot{\gamma}_{\tiny\mbox{eff}}= \frac{3U_{av}}{\delta},
\end{equation}
where $\delta$ is a characteristic lengthscale and the choice of the numerical prefactor $3$ is technically detailed in the Sec.~III of the SM~\cite{suppmat}(see also comments below). For a rectangular microchannel, $\delta$ should be chosen as the smallest size between $H$ and $W$. However, in our study, these are of the same order of magnitude, and the choices $\delta=W$ or $\delta=H$ in Eq.~\eqref{eq:shear_av} are both appropriate to analyze the droplet size as a function of the Capillary number for fixed channel geometry. We choose $\delta=W$.
To highlight the importance of setting $\delta$ to be equal to the smallest scale between $H$ and $W$, one could perform a study by keeping the Capillary number fixed while changing the channel geometry, with situations where $W$ is much different from $H$: this is outside the scope of the present paper, but surely deserves dedicated scrutiny in the future.
Since $U_{av}=Q_c/(W H)$, the value of an effective viscosity $\eta_c (\dot{\gamma_{\tiny\mbox{eff}}})$ can be computed by the power law fluid model (\ref{eq:PowerLawFluidtext}) in terms of $Q_c$. Within the investigated range of $Q_c$, the computed shear rates span from $\approx1$ s$^{-1}$ to $\approx 2 \times 10^3$ s$^{-1}$, which is in the range of validity of the power law model~\cite{macosko_book_94,epjesliding}. In the case of a shear thinning continuous phase, an effective Capillary number $\Cabar$ is then introduced. A balance between viscous thinning effects and pressure gradients (see Sec.~III of the SM) suggests that a proper definition of $\Cabar$ is:
\begin{equation}\label{eq:Cabartext}
\overline {\mbox{Ca}} = n \: \left[ \frac{\eta_c(\dot{\gamma}_{\tiny\mbox{eff}}) U_{av}}{%
\sigma} \right]
\end{equation}
with $n$ being the flow behavior index of the power law fluid (see Eq.~\eqref{eq:PowerLawFluidtext}). In the case of a Newtonian fluid, $\overline {\mbox{Ca}}$ reduces to the usual $\mbox{Ca}$ number defined in Eq.~\eqref{eq:Catext} since $n = 1$ and $\eta_c(\dot{\gamma}_{\tiny\mbox{eff}})$ is a constant. A few comments regarding the definition of the effective Capillary number in Eq.~\eqref{eq:Cabartext} and the effective shear rate in Eq.~\eqref{eq:shear_av} are in order. First, we notice that the effective shear rate $\dot{\gamma}_{\tiny\mbox{eff}}$ is here imposed via the geometrical lengthscale $\delta$ and the average velocity $U_{av}$. This differs from the definition adopted in the analysis of non-Newtonian sliding droplets~\cite{epjesliding,varagnolo17_softmatter}, where a ``phenomenological'' lengthscale was introduced only \emph{a posteriori}, and whose value was derived by imposing that the non-Newtonian data match the Newtonian ones for small Capillary numbers and small driving forces.\\ 
There are also some technical differences in the definition of the effective Capillary number, if compared with the definition adopted in a recent study by Roumpea and coworkers~\cite{roumpea17_aiche}: first, we use the prefactor $3$ in front of the effective shear rate, while the definition in ref.~\cite{roumpea17_aiche} has a refined $n$-dependent prefactor~\cite{Lindner00}, which is in principle more accurate. However, the changes in the prefactor induced by a change in $n$ are small if compared to the changes we have in the average flow velocity, which spans several orders of magnitude. Second, our definition of the effective Capillary number shows a proportionality with the flow behavior index $n$, while the definition in~\cite{roumpea17_aiche} does not. We find this as a consequence of the balancing between pressure forces and viscous dissipation, as explained in Sec.~III of the SM~\cite{suppmat}. Working with power law fluids where $n$ does not change too much, it is somehow difficult to further assess these differences on a more quantitative basis. More work is needed in this direction, possibly including a more ample variety of fluids with many realizations of $n$.

%
\subsection{Numerical Simulations}\label{sec:numerics}
Our numerical simulations rely on Lattice Boltzmann models (LBM). LBM are mesoscopic methods, tracking the evolution of the probability distribution function for particles in a discretized space and time domain. The hydrodynamics of Navier-Stokes (NS) equations is recovered from the coarse-grained behavior of the system~\cite{Succi01}. The mesoscopic nature of the methods can provide many advantages over atomistic approaches, making the LBM especially useful for the simulation of droplets and interfacial dynamics at the microscales. There are various LBM that have been adopted to investigate droplet formation and dynamics in confined T-junction geometries~\cite{van2005droplet,vandersman06,alapati20083d,gong2010lattice,BowerLee11,Gupta09,Gupta10,LiuZhang09b,LiuZhang11,Wuetal,yang2013three,shi2015lattice}. In particular, LBM have already been used for the simulation of non-Newtonian phases in microfluidic T-junctions in Ref.~\cite{shi2015lattice}, where it is shown that the LBM are actually able to capture sizable effects of non-Newtonian rheology on the droplet formation process. The present numerical work follows other studies by some of the authors~\cite{gupta2016effects,gupta2016lattice,SbragagliaGuptaScagliarini,SbragagliaGupta,epjesliding}, where a NS description based on LBM has been coupled to constitutive equations for different polymer dynamics. Specifically, in two recent papers~\cite{gupta2016effects,gupta2016lattice}, three-dimensional (3D) simulations are carried out to quantify the effects of elasticity in the breakup processes in confined T-junctions~\cite{gupta2016effects} and cross-junctions~\cite{gupta2016lattice}, and to investigate the effects of thinning~\cite{epjesliding} in open microfluidic geometries~\cite{Kusumaatmaja,Gabbanelli}. Here the focus is instead on thinning effects in the T-junction geometry. We refer the interested reader to our previous papers where all the relevant LBM technical details are discussed~\cite{gupta2016effects,gupta2016lattice,SbragagliaGuptaScagliarini,epjesliding}. In Sec.~VI A of the SM~\cite{suppmat}, we just briefly recall the relevant NS macroscopic equations that are integrated with the LBM. Some numerical benchmarks for Newtonian liquids are further illustrated and discussed in Sec.~VI B of the SM.
%
%
\section{Results}\label{sec:experimental_results}
In this section we report the systematic results obtained for the droplet breakup in microfluidic T-shaped junctions. Section~\ref{sec:size} shows the variation of the droplets size in the various regimes up to jetting, either for Newtonian or shear thinning continuous phases. The identification of the breakup regimes is obtained by constructing comprehensive breakup maps for all the liquids (see Sec.~IV of the SM). The rescaled data of all liquids in terms of the effective Capillary number are presented in Sec.~\ref{sec:rescaling}. This analysis is suitably complemented with quantitative results based on numerical simulations of purely thinning fluids in the squeezing-to-dripping transition. Finally, based on numerical simulations, we show and analyze in Sec.~\ref{sec:numerical_results} the velocity and stress profiles developed along the channel. Time evolution of the stresses is provided in the movies of the SM.

\subsection{Droplet size}\label{sec:size}
\begin{figure*}[h]
  \includegraphics[width = 0.93\textwidth]{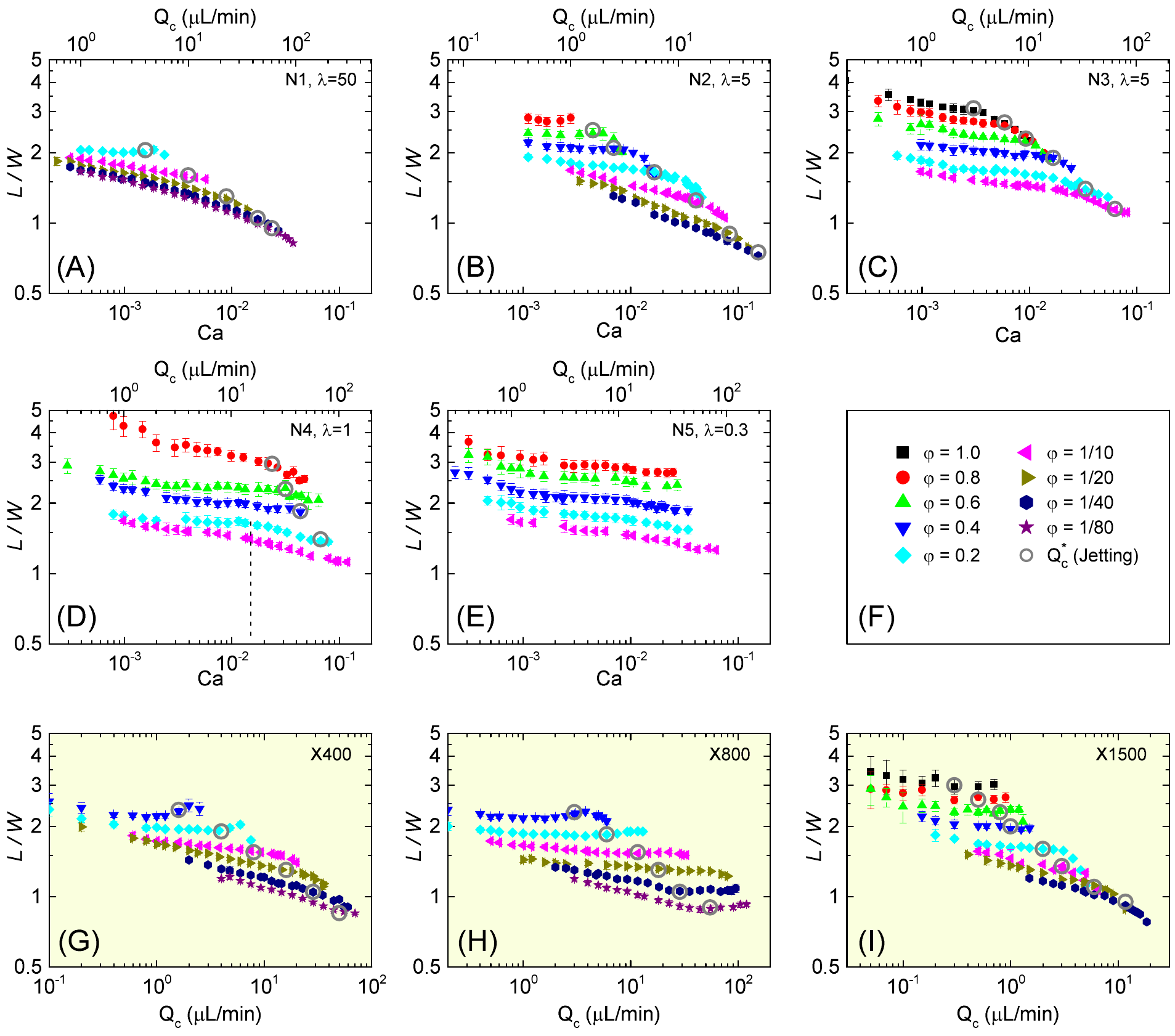}
\caption{Normalized, dimensionless length $L/W$ of the droplets formed at the microfluidic T-junction as a function of either the flow rate of the continuous phase $Q_c$ [top axis of panels (A)-(E), bottom axis of  panels (G)-(I)] or the corresponding Capillary number $\mathrm{Ca}$ [bottom axis in panels (A)-(E)] for different values of the flow rate ratio $\varphi$, indicated by different symbols according to the legend reported in panel (F). Panels (A)-(E) refer to the Newtonian fluids listed in Table~\ref{tbl:expNewton}, while panels (G)-(I) correspond to the polymers listed in Table~\ref{tbl:expThinning}. Open circles mark the flow rate $Q_c^*$ corresponding to the onset of the jetting regime, accordingly to the maps shown in Fig.~S1 of the SM.
}
\label{fig:fig2ExpData}
\end{figure*}

In the presence of a shear thinning continuous phase, we can still identify the breakup regimes commonly reported with Newtonian liquids~\cite{Seeman12,baroud_dynamics_2010,Demenech07} shown in the left snapshots of Fig.~\ref{fig:exp_snaps} at increasing $\Ca$. In Fig.~\ref{fig:exp_snaps}(A), droplets form at the junction and fill the channel assuming a plug shape. The dispersed phase completely obstructs the channel, leading to an increase in the pressure upstream, which eventually breaks-up the interface into a droplet. In this ``squeezing'' regime the droplet size does not strongly depend on Ca, but only on the flow rates~\cite{Garstecki06,Demenech07,christopher_experimental_2008,LiuZhang09}. In Fig.~\ref{fig:exp_snaps}(C) the droplets are emitted before they can block the channel and their formation is due to the action of the viscous shear stress.  In this ``dripping'' regime, the droplet size decreases with $\Ca$~\cite{thorsen_dynamic_2001,Demenech07,christopher_experimental_2008}. Finally, Fig.~\ref{fig:exp_snaps}(E) shows that, at high $\Ca$, the detachment point moves progressively downstream and the breakup process signals the emergence of a ``jetting'' mode~\cite{Demenech07}.  The snapshots in Fig.~\ref{fig:exp_snaps}(B) and Fig.~\ref{fig:exp_snaps}(D) show the intermediate cases right before the transition to dripping and jetting, respectively. The right snapshots of Fig.~\ref{fig:exp_snaps} refer to oil drops in Xanthan solutions. They clearly show that, by varying the flow rate $Q_{c}$ of the continuous phase, it is possible to reproduce the same regimes reported for the Newtonian systems. 
%
%
Figure~\ref{fig:fig2ExpData} shows the dependence of the normalized droplet length $L/W$ as a function of the flow rate $Q_c$ of the continuous phase (for Xanthan solutions) or both $Q_c$ and the Capillary number $\Ca$ (for Newtonian liquids) at different flow rate ratios $\varphi=Q_d/Q_c$ and $\lambda$. Figures~\ref{fig:exp_snaps}(A)-\ref{fig:exp_snaps}(E) refer to the Newtonian liquids whose details are listed in Table~\ref{tbl:expNewton}, while Figs.~\ref{fig:exp_snaps}(G)-\ref{fig:exp_snaps}(I) correspond to the drops formed in shear thinning continuous phases described in Table~\ref{tbl:expThinning}. These measurements provide a significant addition to the existing literature that is mainly focused on the dependence of the droplet size when $\lambda \leq 1 $ and $\varphi < 1$~\cite{christopher_experimental_2008,Demenech07,LiuZhang09}. For each $\varphi$, the data stops at a certain value $Q_c$, which is found to increase as $\varphi$ gets smaller at fixed $\lambda$. It represents either the maximum $Q_c$ for which droplet breakup still occurs before the detection window of Fig.~\ref{fig:exp_snaps}-A, or the maximum $Q_c$ reachable by our setup. 
The open circles appearing in the curves mark the flow rate $Q_c^*$ corresponding to the onset of the jetting regime, according to the breakup maps shown in Fig.~S1 of the SM. 
The transition to jetting displays approximately a power law behavior with $\varphi \sim {Q_c}^{-1}$, corresponding to a constant value of $Q_d$, as reported in the literature~\cite{FuTaoTao2015}. Further, as other authors report~\cite{langmuir2010}, hysteresis has been observed in the transition to jetting, i.e. after reaching the full jetting regime [Fig.~\ref{fig:exp_snaps}(E)] by increasing $Q_c$, the jet persists even after decreasing $Q_c$.

The size of droplets transported by a Newtonian liquid is observed to increase at increasing $\varphi$, and decrease at increasing $\Ca$~\cite{Garstecki06,christopher_experimental_2008}. This trend is consistent with previous studies performed in similar conditions~\cite{christopher_experimental_2008,LiuZhang09}.
For instance, at small Ca, the droplet size is independent of the viscosity ratio [see, for instance, the data at $\varphi=0.4$ in Fig.~\ref{fig:fig2ExpData}(B)-\ref{fig:fig2ExpData}(E)] and increases with the flow rate ratio [compare, for instance, the data at $\varphi=0.4$ and $\varphi=0.6$ in Fig.~\ref{fig:fig2ExpData}(B)-\ref{fig:fig2ExpData}(E)], in agreement with the scaling argument valid for the squeezing regime~\cite{Garstecki06}. Furthermore, the change in the slope found at $\Ca \simeq 10^{-2}$ for the curve $\varphi=0.2$ and $\lambda=1$ in Fig.~\ref{fig:fig2ExpData}-D  (see dashed line) is very close to the transition from squeezing to dripping calculated in phase-field numerical simulations of immiscible fluids at $\Ca \approx 0.015$ for $\varphi=0.25$ and $\lambda=1$~\cite{Demenech07}. We point out that, with our geometry, the dripping regime is not very pronounced because the widths of the two inlet channels are the same~\cite{LiuZhang09b,christopher_microfluidic_2007}.
%
%
%
%
%
\\
The size of droplets carried by Xanthan solutions is observed to increase with $\varphi$ and initially decrease with $Q_c$, just like the Newtonian counterpart. However, at high $Q_c$ the droplets become longer in the direction of the flow. The analysis of their volume, performed by measuring the frequency of the droplets production as done in Ref.~\cite{christopher_experimental_2008} reveals that the volume of the droplets actually decreases with increasing $Q_c$ (data not shown).
The elongation of the droplets carried by a shear thinning phase can be seen in Fig.~\ref{fig:exp_snaps}-H and it has also been recently reported in a  similar study~\cite{roumpea17_aiche}.
Particle image velocimetry measurements performed in Ref.~\cite{roumpea17_aiche} show that the film thickness of the continuous phase surrounding the droplets increases with increasing concentration of Xanthan solution. More precisely,  for a Newtonian system, the liquid film corresponds to approximately $3\%$ of
the channel diameter, whereas for the 2000 ppm Xanthan gum system the film thickness is almost $10\%$ of the channel diameter~\cite{roumpea17_aiche}.
It is noteworthy that elongation takes place when the shear forces appear to be quite consistent, and noticeably in the emergence of the jetting regime. 
Apart from this elongation, the droplet production in shear thinning continuous phases is qualitatively similar to that occurring in purely Newtonian systems. 
%
%
\subsection{Rescaling over $\Cabar$}\label{sec:rescaling}
\begin{figure*}[hp]
\begin{center}
\subfigure{\includegraphics[scale=0.9,trim=0 -20 0 0,clip]{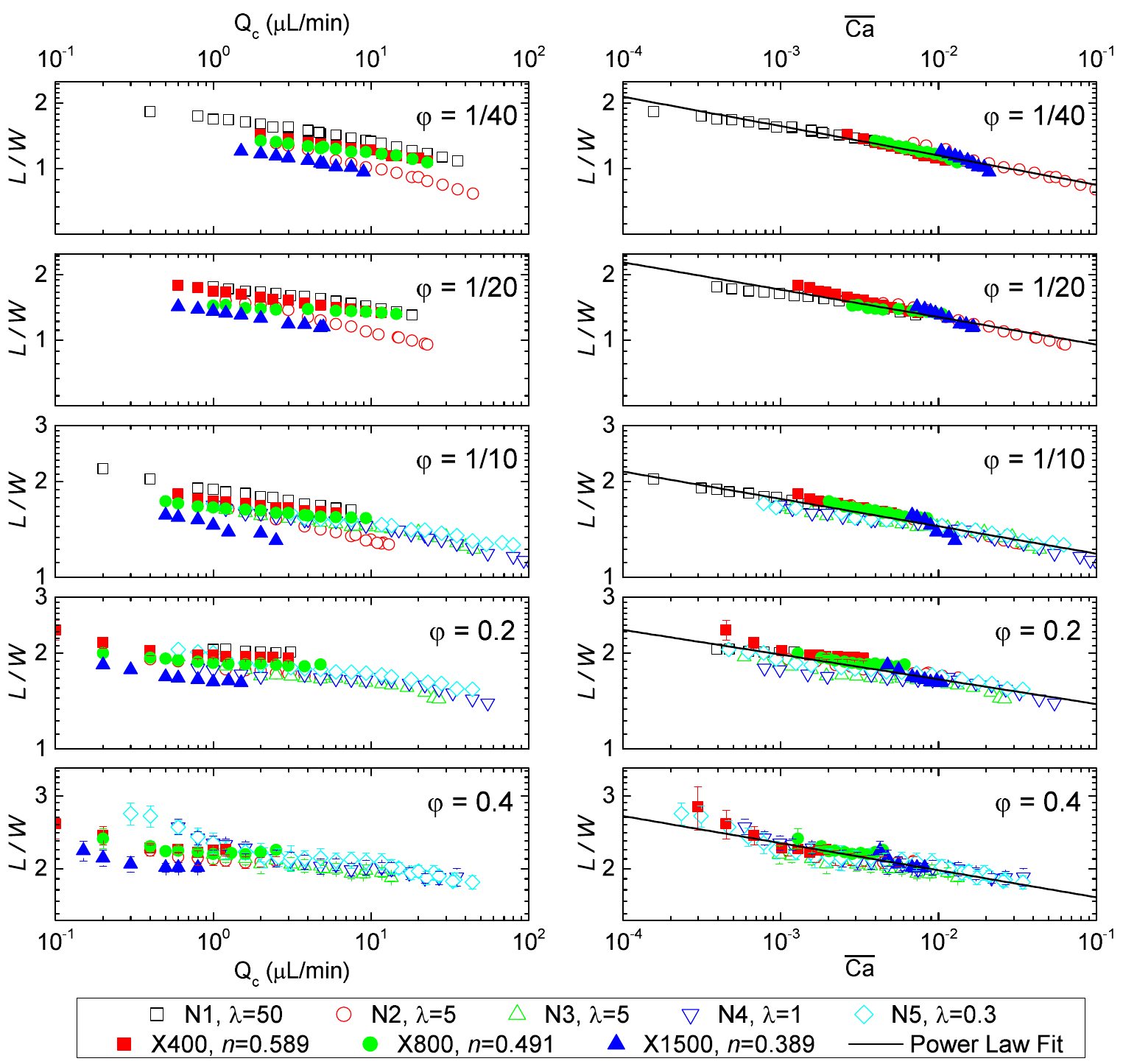}}\\
\subfigure{\includegraphics[scale=0.9]{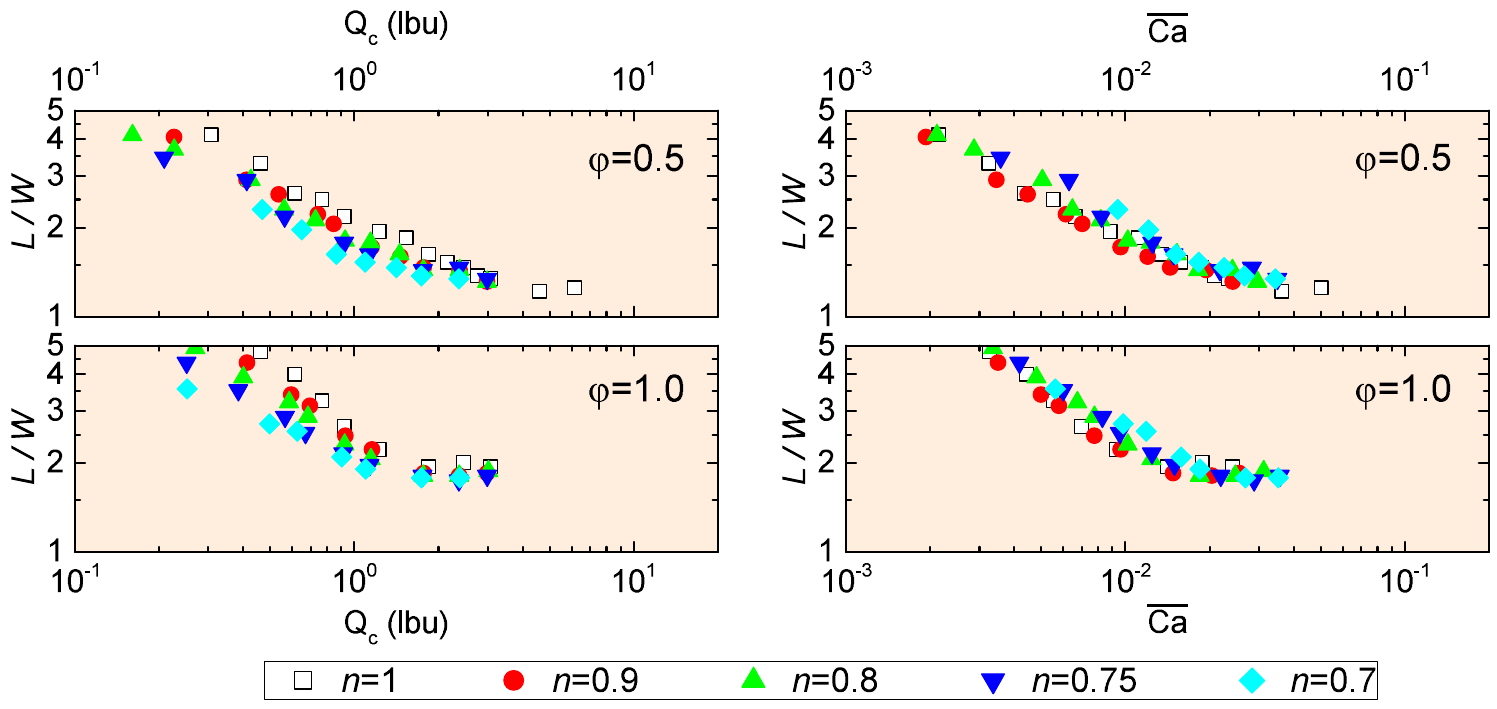}}
\caption{Experimental (top, white panels) and numerical (bottom, orange panels) normalized droplet length $L/W$ as a function of the flow rate of the continuous phase $Q_c$ (left column) and the effective Capillary number $\Cabar$ defined in Eq.~\eqref{eq:Cabartext} (right column), for different values of the flow rate ratio $\varphi$, which is increasing from top to bottom. All the graphs are in log-log scale. Open symbols refer to Newtonian fluids, filled symbols of the experimental graphs  to shear thinning Xanthan solutions at different concentrations, full symbols of the numerical graphs to power law fluids with different flow behavior indexes $n$. The line is a power law fit to the experimental data, with fitting parameters reported in Table~\ref{tbl:powlawfitLength}.}
\label{fig:rescale_L}
\end{center}
\end{figure*}

A quantitative comparison between Newtonian and non-Newtonian systems is then performed for selected values of the viscosity ratio $\lambda$,  and of the flow rate ratio $\varphi$.
In the left column of Fig.~\ref{fig:rescale_L} the average normalized droplet length $L/W$ is reported as a function of the continuous flow rate $Q_c$ for chosen $\varphi$, covering  more than one decade, from $\varphi = 1/40$ to $\varphi = 0.4$.
Data taken at $Q_c \lesssim 0.1~\mu L/ \textrm{min}$ are discarded due to poor reproducibility of our syringe pumps in this range. Similarly, droplets formed in the jetting regime are not considered because of elongation effects. 
The graphs clearly show that the droplets formed in the shear thinning Xanthan solutions get smaller as the Xanthan concentration is increased, at a given $\varphi$.
In addition, the size difference between droplets carried by Xanthan at 1500 ppm and the ones formed in water is about $30\%$ for about one decade of $\varphi$.
However, when plotted against $\Cabar$, a remarkable degree of collapse on the same curve is observed for all the data. With such rescaling, the droplets produced at a given $\varphi$ display essentially the same size either they are formed in Newtonian or non-Newtonian, shear thinning, continuous phases.
The droplet size is found to scale as a power law $L/W \sim A~ \Cabar~^{\beta}$, with $A$ being a function of $\varphi$~\cite{christopher_experimental_2008,Tan,chiarello2015generation}. The straight lines of Fig.~\ref{fig:rescale_L} are fits to the data according to this power law and the resulting parameters are reported in Table~\ref{tbl:powlawfitLength} for each value of $\varphi$. Both $A$ and $\beta$ are found to increase with $\varphi$. In particular, $A \sim \varphi~^{(0.32 \pm 0.02)}$, the exponent being similar to those reported by Refs.~\cite{Tan,chiarello2015generation}.
These observations provide a direct, experimental evidence of the validity of the $\Cabar$ number, Eqs.~\eqref{eq:shear_av} and~\eqref{eq:Cabartext}, to properly capture the shear distribution inside the microfluidic channel, when a shear thinning fluid is flowing at a given $Q_{c}$.
%
%
%
%
%
\begin{table*}[t!]
  \small  
   \begin{tabular*}{0.4\textwidth}{@{\extracolsep{\fill}} l l l }
    \hline
    $\mathbf{\varphi}$ & $\mathbf{\textit{A}}$ & $\mathbf{\beta}$  \\
    \hline  


0.025 & $  0.62 \pm  0.01 $ & $ -0.135  \pm   0.003$  \\
0.05 &  $ 0.71  \pm  0.02 $ & $  -0.126 \pm   0.005$  \\
0.1 &  $ 0.97  \pm   0.01 $ & $ -0.086  \pm   0.002$  \\
0.2 &  $ 1.16  \pm   0.03 $ & $ -0.078  \pm   0.005$  \\
0.4 &  $ 1.47  \pm   0.02 $ & $ -0.065  \pm   0.003$  \\

    \hline
   \end{tabular*}
   \caption{Fitting parameters $A$ and $\beta$ used to fit the experimental value of the normalized droplet length $L/W$ accordingly to $L/W=A~(\Cabar)^{\beta}$ corresponding to the values of the flow rate ratio $\varphi$ showed in Fig.~\ref{fig:rescale_L}. (white back ground, right panels).} 
\label{tbl:powlawfitLength}
\end{table*}
%
%
\subsection{Velocity and stress distribution in the channel}\label{sec:numerical_results}
To better understand the role played by the Xanthan solutions, the droplet breakup experiments are complemented with realistic numerical simulations of purely thinning fluids, which allow us to directly visualize the viscous stress and velocity profiles inside the microfluidic channels and to highlight the contribution of shear thinning. Indeed, especially at high concentrations, Xanthan solutions may be affected by normal stress effects~\cite{epjesliding,whitcomb78Xanth} (see also SM, Sec~V), whereas the numerical simulations here performed do not include such effects.
Figures~\ref{fig:Qc_choreo_PROFILE} and \ref{fig:Qc_choreo} report snapshots of the droplet formation process in the squeezing-to-dripping transition, for a representative case with fixed $Q_c = 1.23$ lbu (lattice Boltzmann units) and flow rate ratio $\varphi = 0.5$. Both Newtonian [$n=1$, \label{fig:Qc_choreo_PROFILE} 4(a) and 4(b)] and non-Newtonian [$n=0.9$, \label{fig:Qc_choreo_PROFILE} 4(c) and 4(d); $n=0.75$, \label{fig:Qc_choreo_PROFILE} 4(e) and 4(f)] cases are analyzed. Specifically, in Fig.~\ref{fig:Qc_choreo_PROFILE}, the density contours of the dispersed phase are overlaid on the velocity vector field evaluated at half of the channel height. For a fixed $Q_c$, the snapshots of velocity vectors share qualitative features, being more intense in the center of the channel. Using the velocity profiles, the viscous stress in the continuous phases is also computed. This is overlaid on the density contours in Fig.~\ref{fig:Qc_choreo}. 
Overall, we observe that by decreasing the power law index $n$ the film thickness between the front of the formed droplet and the walls slightly increases and droplets have the tendency to develop a ``bullet-shaped'' profile, which recalls the experimental finding in Fig.~\ref{fig:exp_snaps}-H and the experimental observations in Ref.~\cite{roumpea17_aiche}. Moreover, for the non-Newtonian fluids, the droplet size comes out to be smaller in comparison to the Newtonian case, and the effect is more pronounced at smaller thinning exponents. This well echoes the experimental findings reported in the left panel of Fig.~\ref{fig:rescale_L}. We actually see in Fig.~\ref{fig:Qc_choreo} that the viscous stress is more intense close to the boundaries at decreasing $n$. This explains the observed discrepancies in droplet sizes, in that whenever a larger viscous stress is present close to the wall, this results in a smaller droplet size. We point out that the use of thinning exponents smaller than $n \approx 0.7$ faces a problem of numerical instabilities associated with the multicomponent LBM used, since the viscosity ratio achieved in different regions of the same numerical simulation is relatively large~\cite{PREnostro}. Hence, the range of thinning exponents differs from the experimental counterpart explored in Fig.~\ref{fig:rescale_L}; however, numerical simulations are still useful to highlight to what degree the observed results originate from a combination of hydrodynamics and bulk thinning phases without influence of elastic effects~\cite{epjesliding,whitcomb78Xanth}. This is done in the bottom (orange) panels of Fig.~\ref{fig:rescale_L}. Although the functional dependency between the droplet size and Capillary number looks a bit more pronounced for the $\varphi$ analyzed in the numerics, the overall picture emerging from Fig.~\ref{fig:rescale_L} highlights that the proposed rescaling arguments work well for a wide range of thinning fluids ($n=0.5-1.0$) and flow rate ratios ($\varphi=1/40-1.0$).
Whenever weak discrepancies emerge, they are of the same order of magnitude for both experiments and numerical simulations; since numerical simulations are performed with purely thinning fluids, it is unlikely to attribute such discrepancies to the weak normal stresses of Xanthan~\cite{epjesliding,whitcomb78Xanth}.
%
%
\begin{figure*}[th!]
\begin{center}
\subfigure[~{\scriptsize $t = t_0 + 9.75 \tshear; n = 1.0$}]{\includegraphics[trim=10 410 0 0, clip, width = 0.4\linewidth]{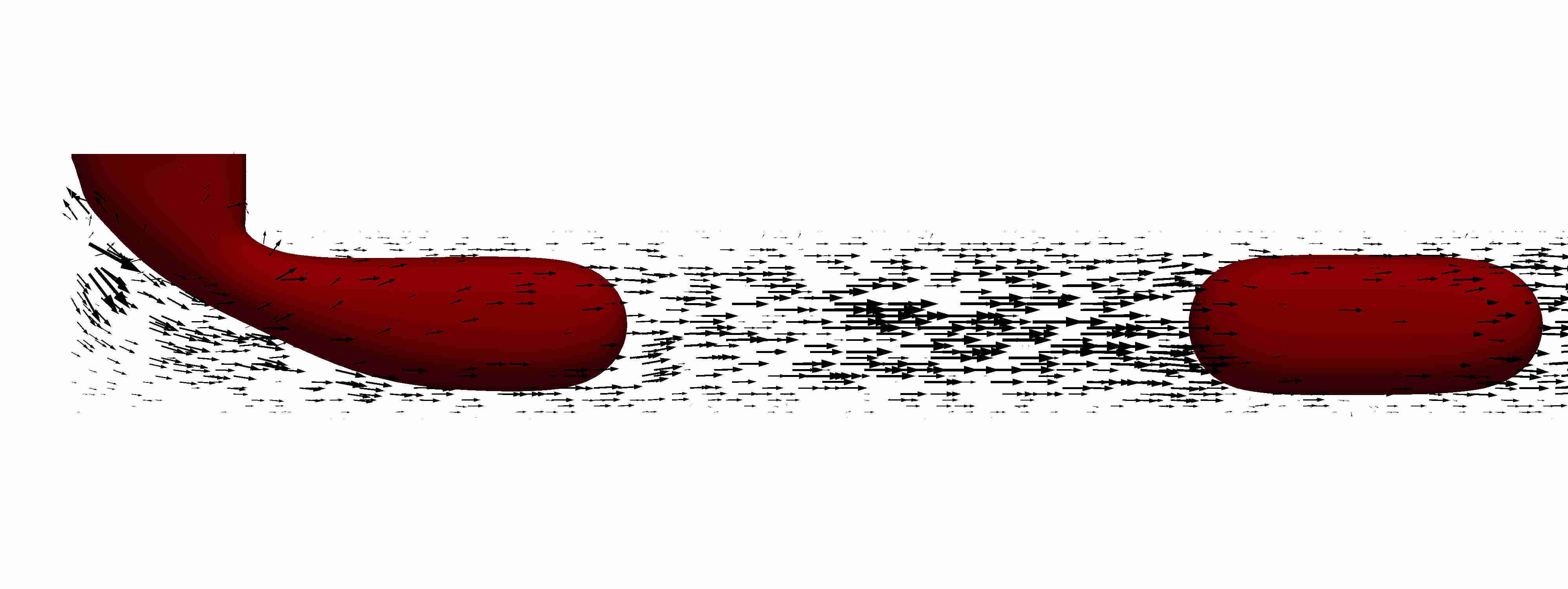}}
\subfigure[~{\scriptsize $t = t_0 + 10.50 \tshear; n = 1.0$}]{\includegraphics[trim=10 410 0 0, clip, width = 0.4\linewidth]{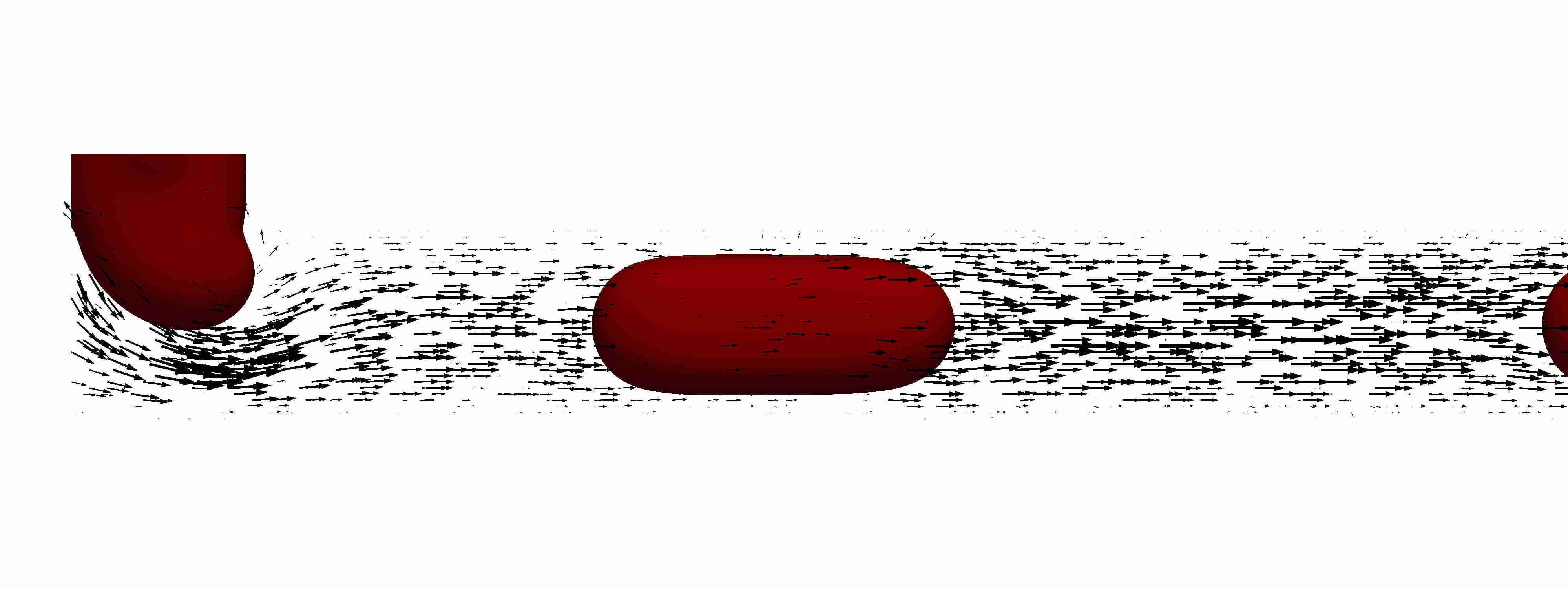}}\\
\vspace{-0.35cm}
\subfigure[~{\scriptsize $t = t_0 + 11.25 \tshear; n = 0.9$}]{\includegraphics[trim=10 410 0 200, clip, width = 0.4\linewidth]{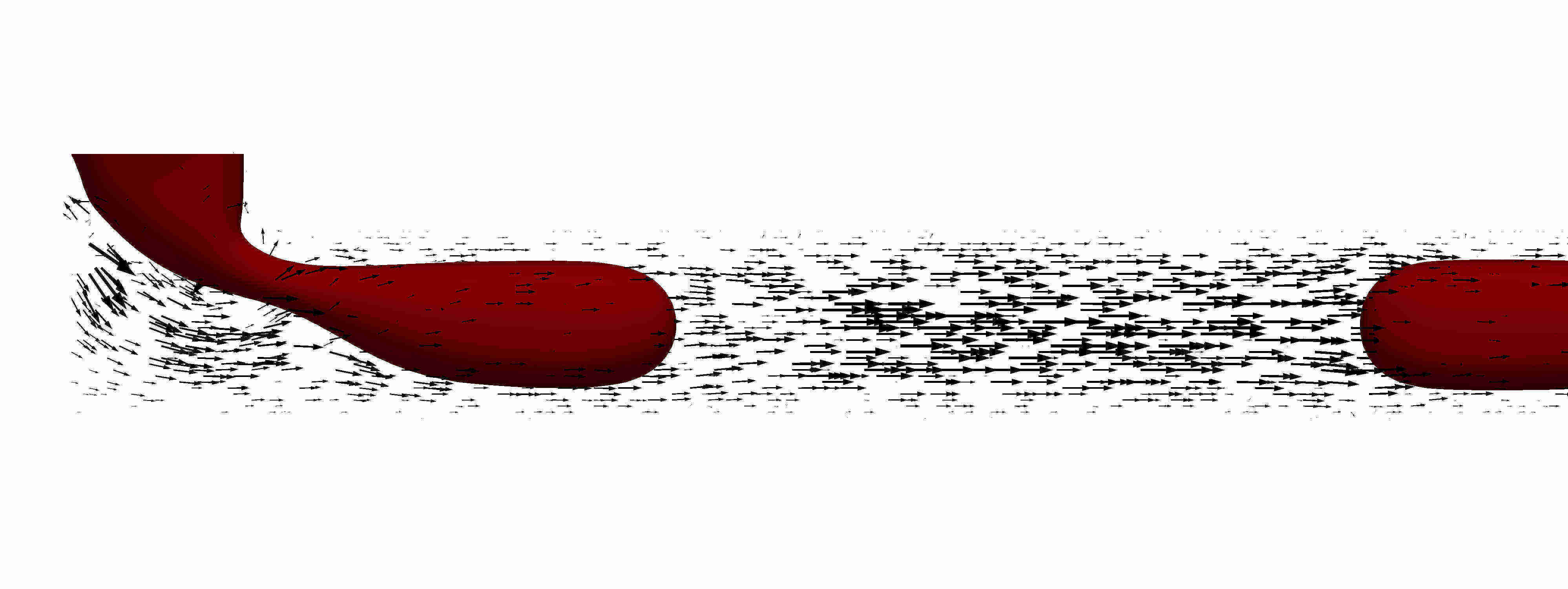}}
\subfigure[~{\scriptsize $t = t_0 + 12.00 \tshear; n = 0.9$}]{\includegraphics[trim=10 410 0 200, clip, width = 0.4\linewidth]{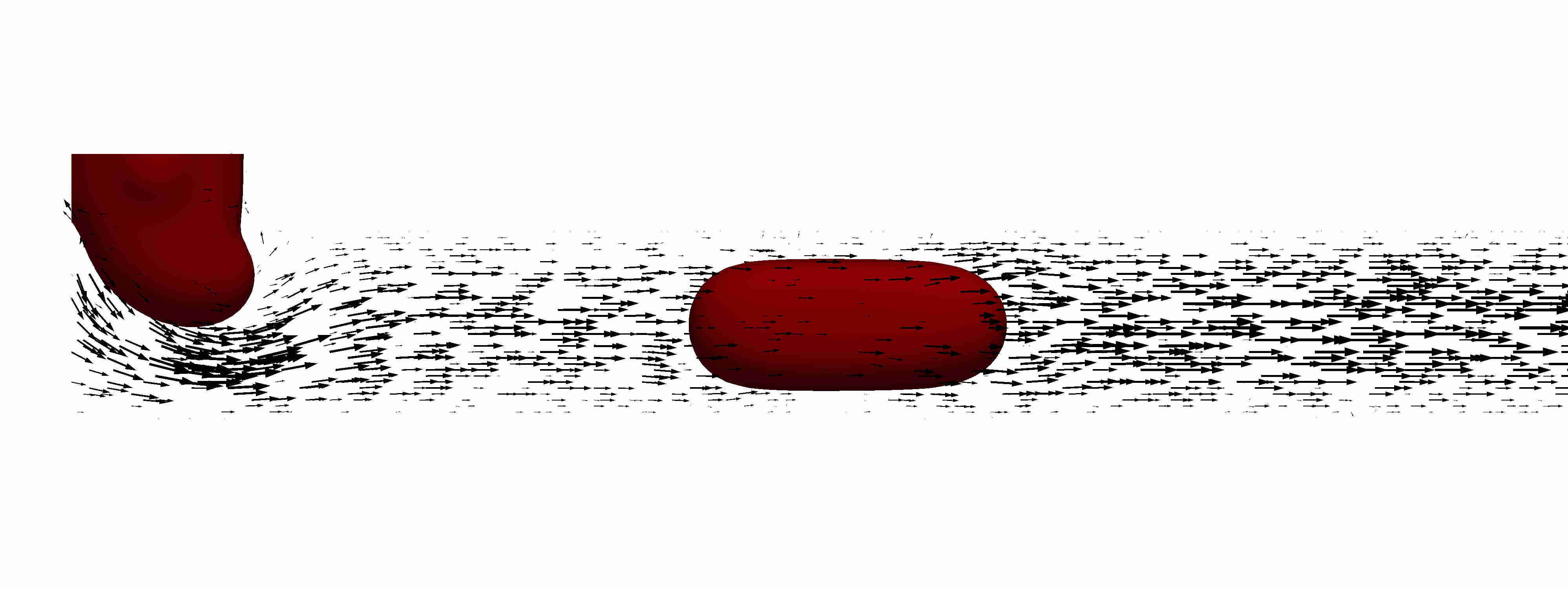}}\\
\vspace{-0.35cm}
\subfigure[~{\scriptsize $t = t_0 + 10.50 \tshear; n = 0.75$}]{\includegraphics[trim=10 410 0 200, clip, width = 0.4\linewidth]{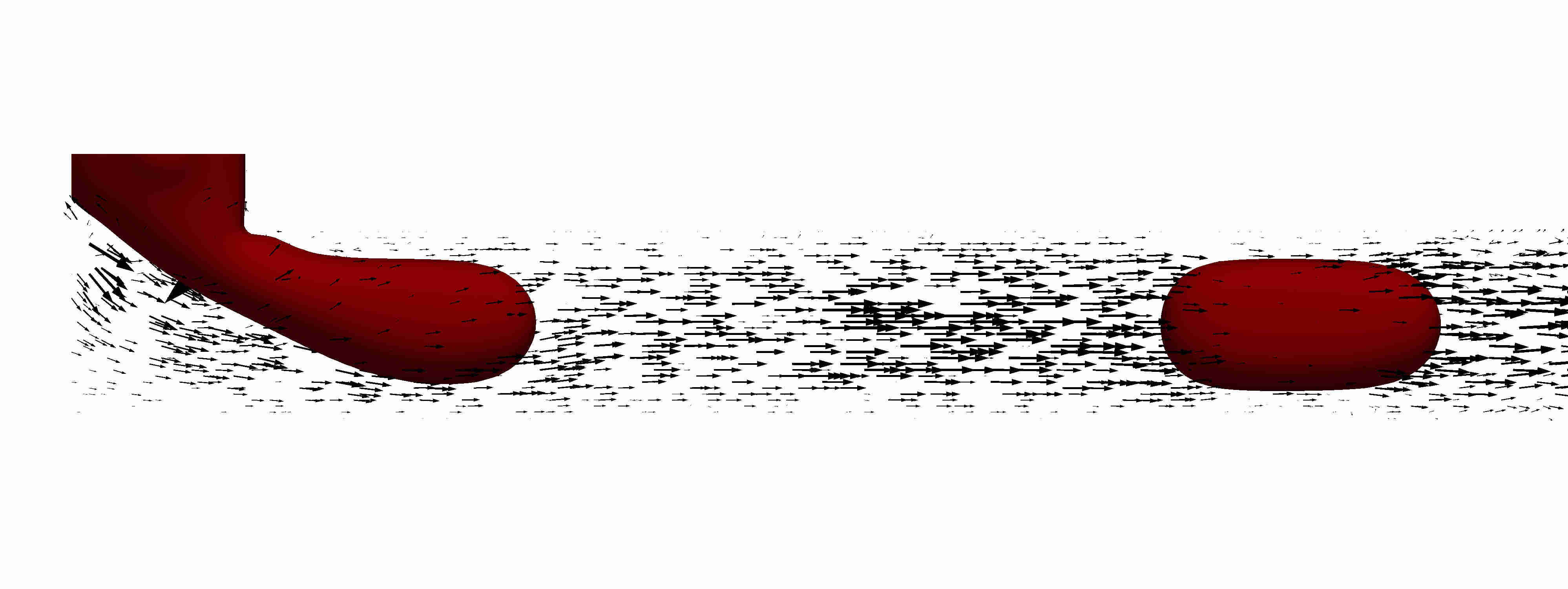}}
\subfigure[~{\scriptsize $t = t_0 + 11.25 \tshear; n = 0.75$}]{\includegraphics[trim=10 410 0 200, clip, width = 0.4\linewidth]{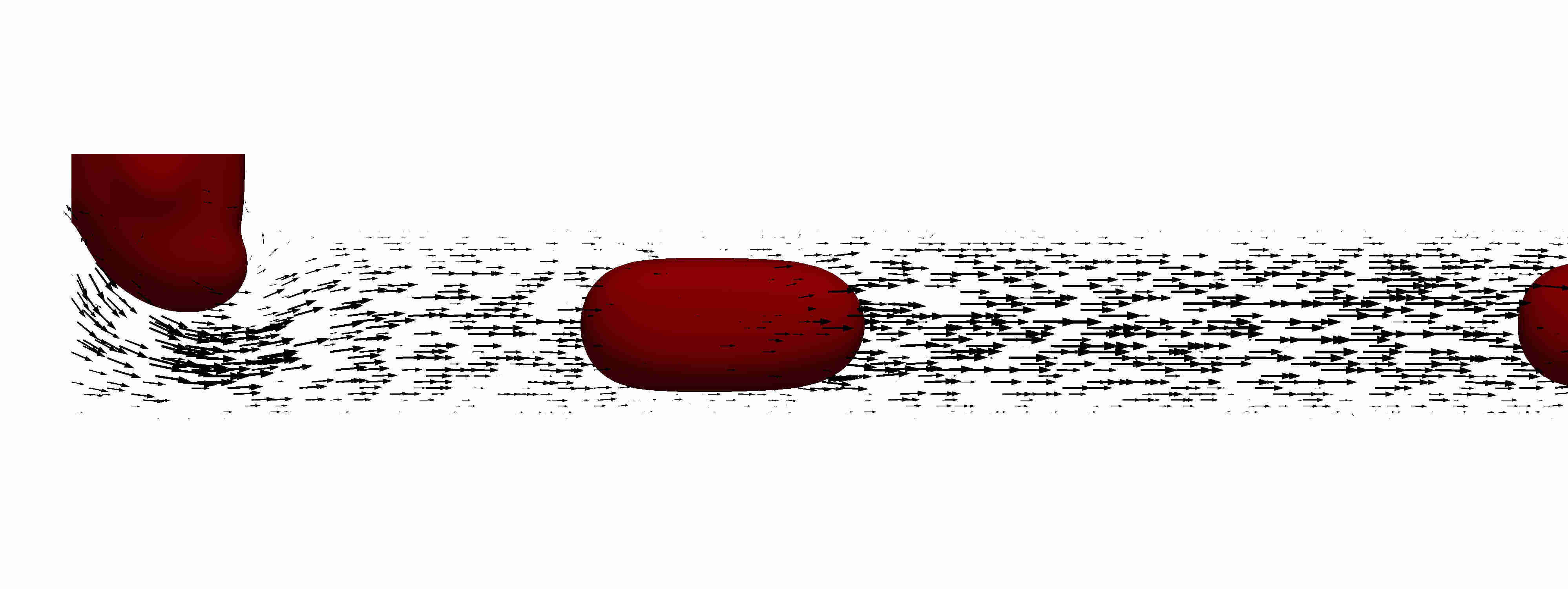}}\\
\caption{Snapshots of the droplet formation process in numerical simulations with LBM, reporting two representative situations before (left column) and after (right column) the breakup process has occurred. 3D snapshots are overlaid on the velocity vector field evaluated on a slice located at half of the channel height. The flow rate ratio is kept fixed to $\varphi = 0.5$ and the flow rate in the continuous phase to $Q_c = 1.23$ lbu. Different power law exponents are used: Newtonian fluid [$n=1$, panels (a-b)], thinning fluid with thinning exponent $n = 0.9$ [panels (c-d)] and $n = 0.75$ [panels (e-f)]. The corresponding stress is reported in Fig.~\ref{fig:Qc_choreo}. Time is made dimensionless using the shear time $\tshear=W/U_{av}$.\label{fig:Qc_choreo_PROFILE}}
\end{center}
\end{figure*}
%
\begin{figure*}[th!]
\begin{center}
\subfigure[~{\scriptsize $t = t_0 + 9.75 \tshear; n = 1.0$}]{\includegraphics[trim=10 410 0 0, clip, width = 0.4\linewidth]{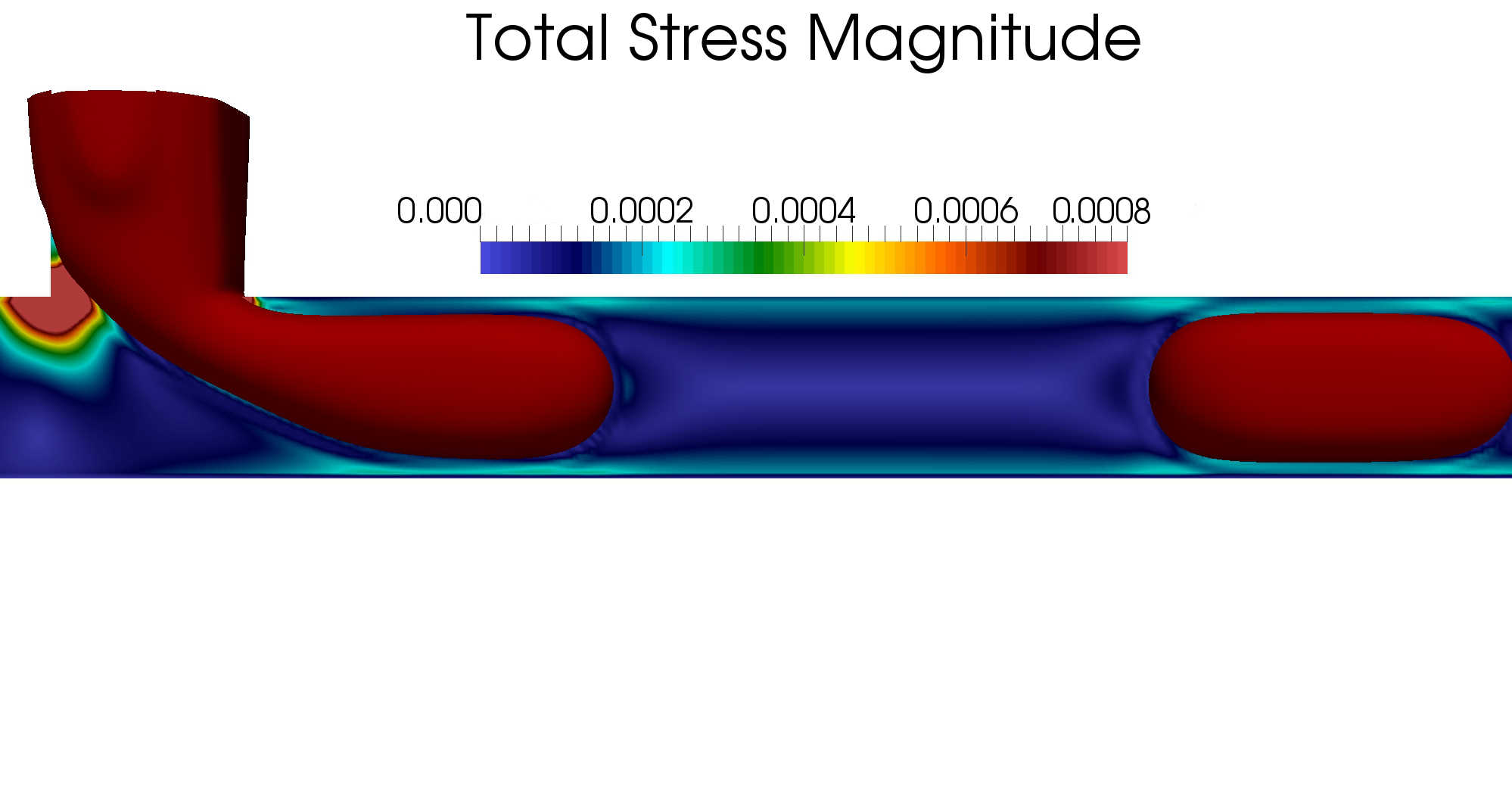}}
\subfigure[~{\scriptsize $t = t_0 + 10.50 \tshear; n = 1.0$}]{\includegraphics[trim=10 410 0 0, clip, width = 0.4\linewidth]{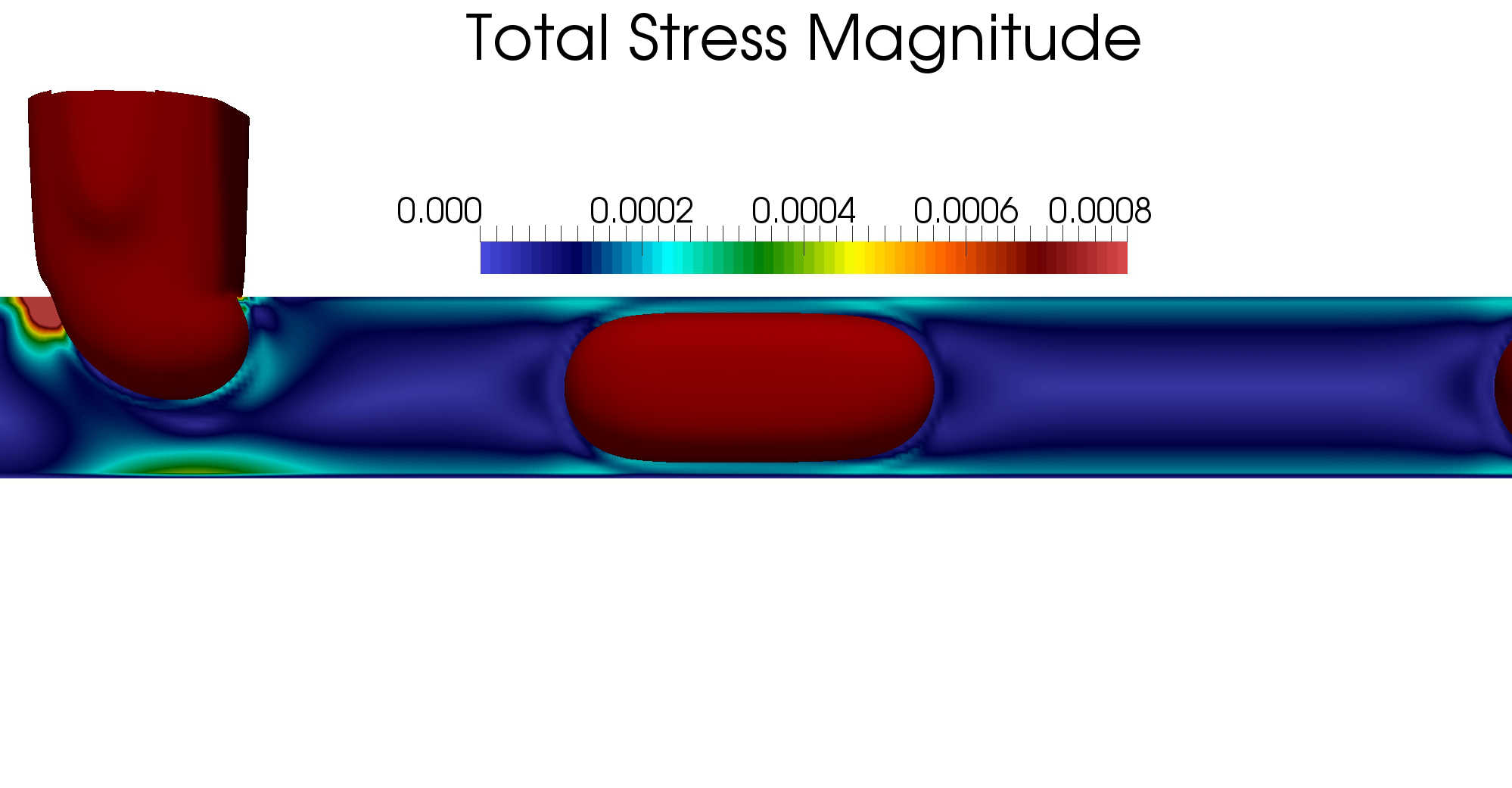}}\\
\vspace{-0.35cm}
\subfigure[~{\scriptsize $t = t_0 + 11.25 \tshear; n = 0.9$}]{\includegraphics[trim=10 410 0 200, clip, width = 0.4\linewidth]{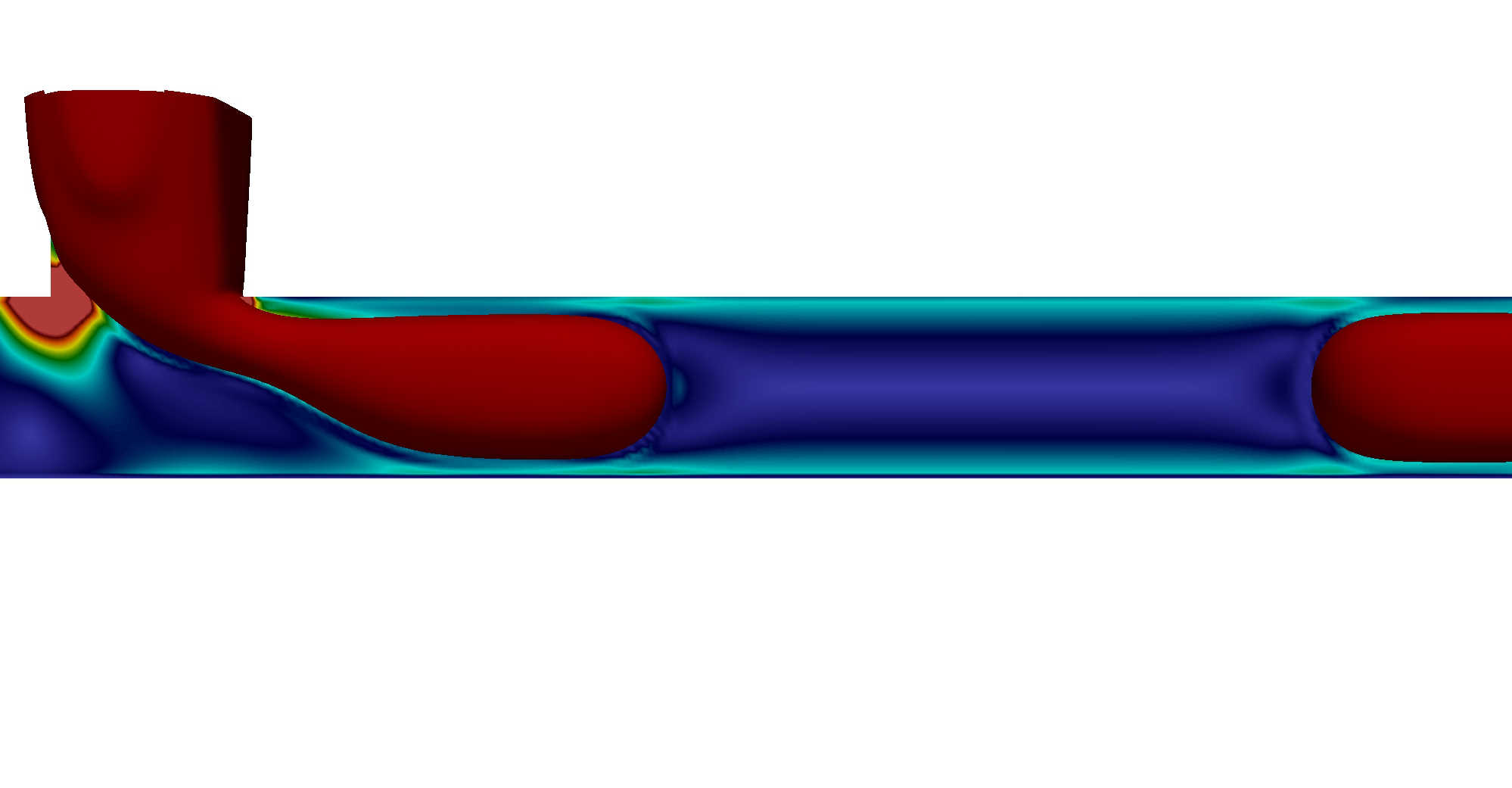}}
\subfigure[~{\scriptsize $t = t_0 + 12.00 \tshear; n = 0.9$}]{\includegraphics[trim=10 410 0 200, clip, width = 0.4\linewidth]{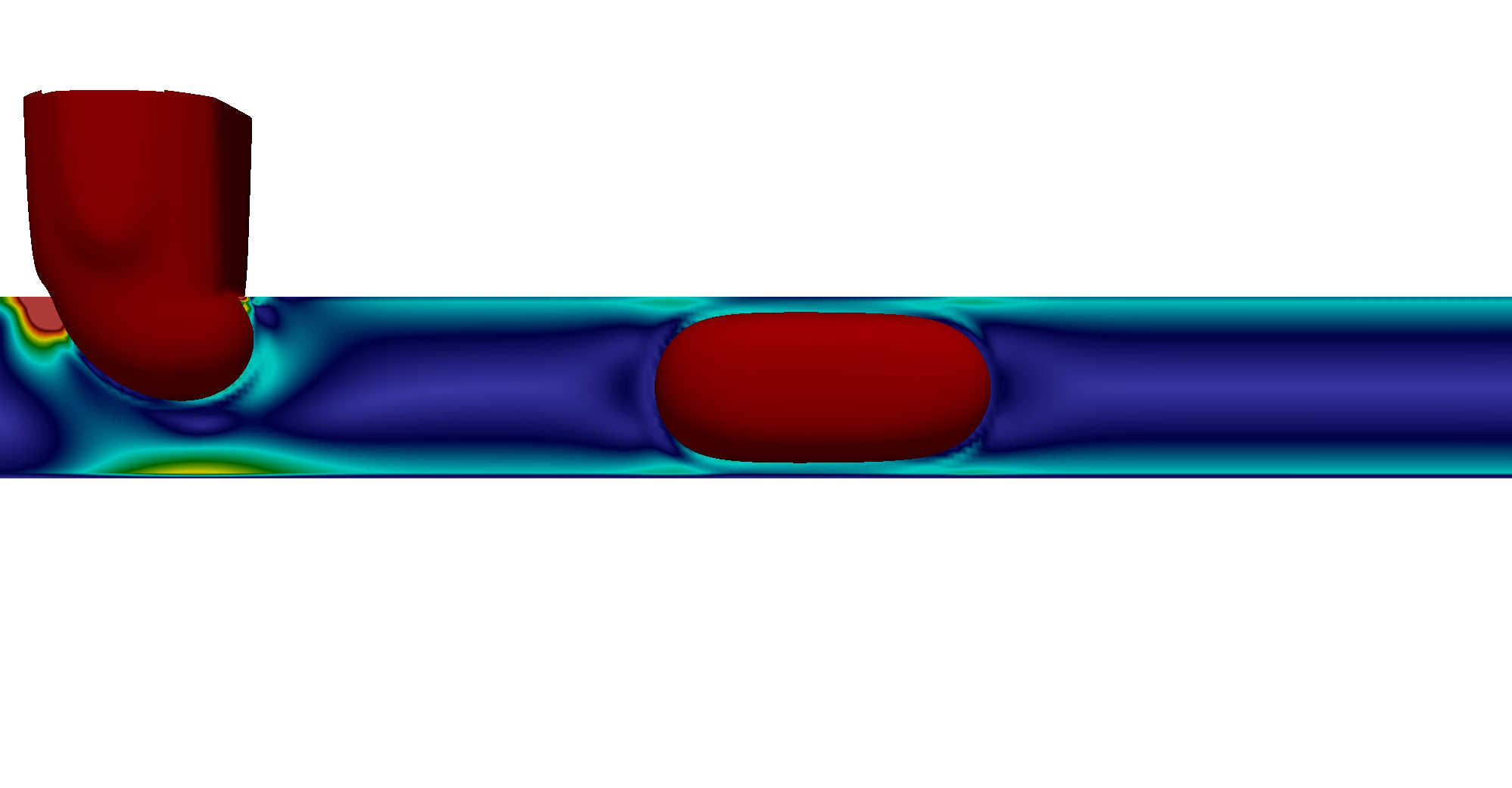}}\\
\vspace{-0.35cm}
\subfigure[~{\scriptsize $t = t_0 + 10.50 \tshear; n = 0.75$}]{\includegraphics[trim=10 410 0 200, clip, width = 0.4\linewidth]{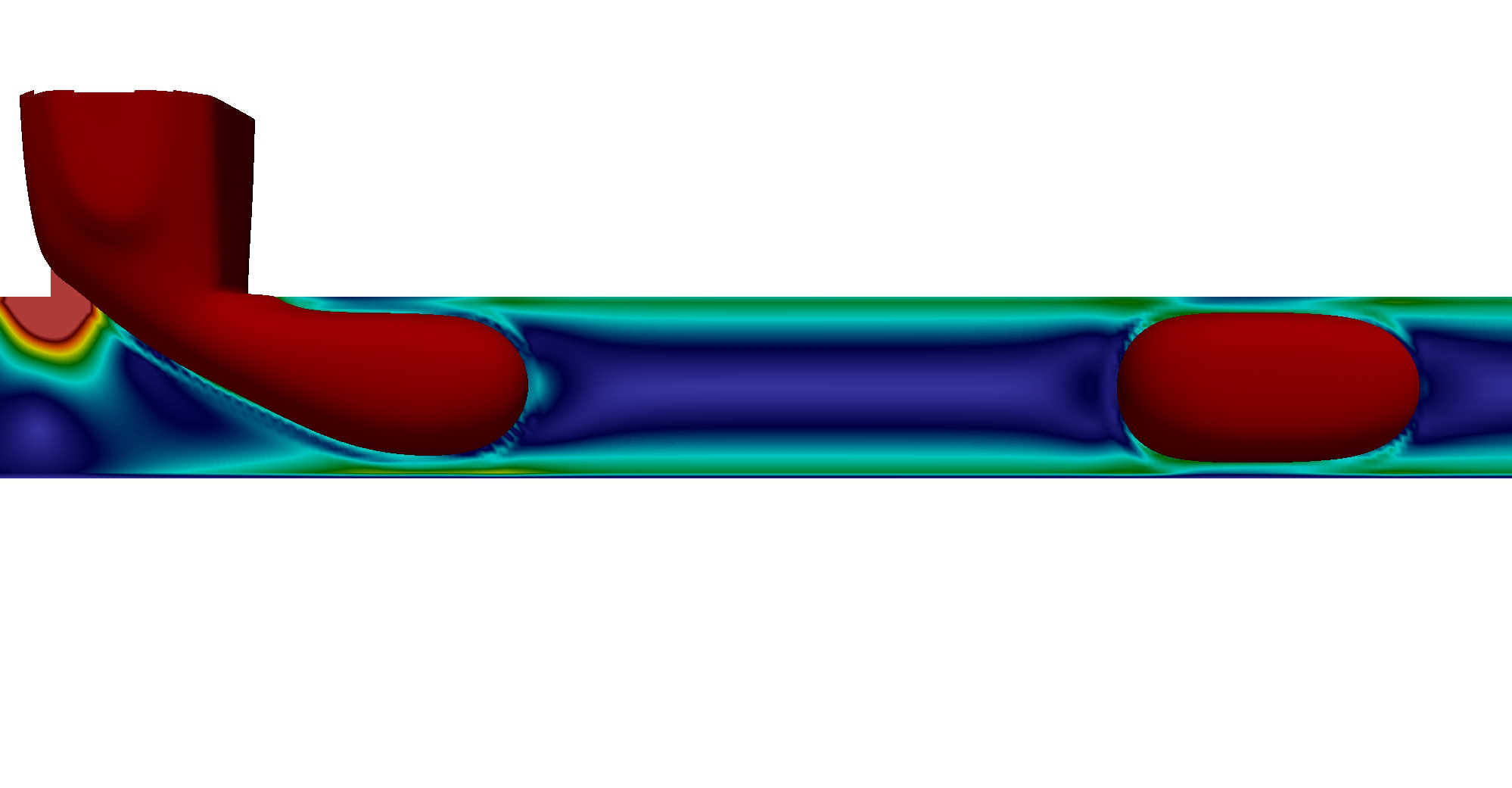}}
\subfigure[~{\scriptsize $t = t_0 + 11.25 \tshear; n = 0.75$}]{\includegraphics[trim=10 410 0 200, clip, width = 0.4\linewidth]{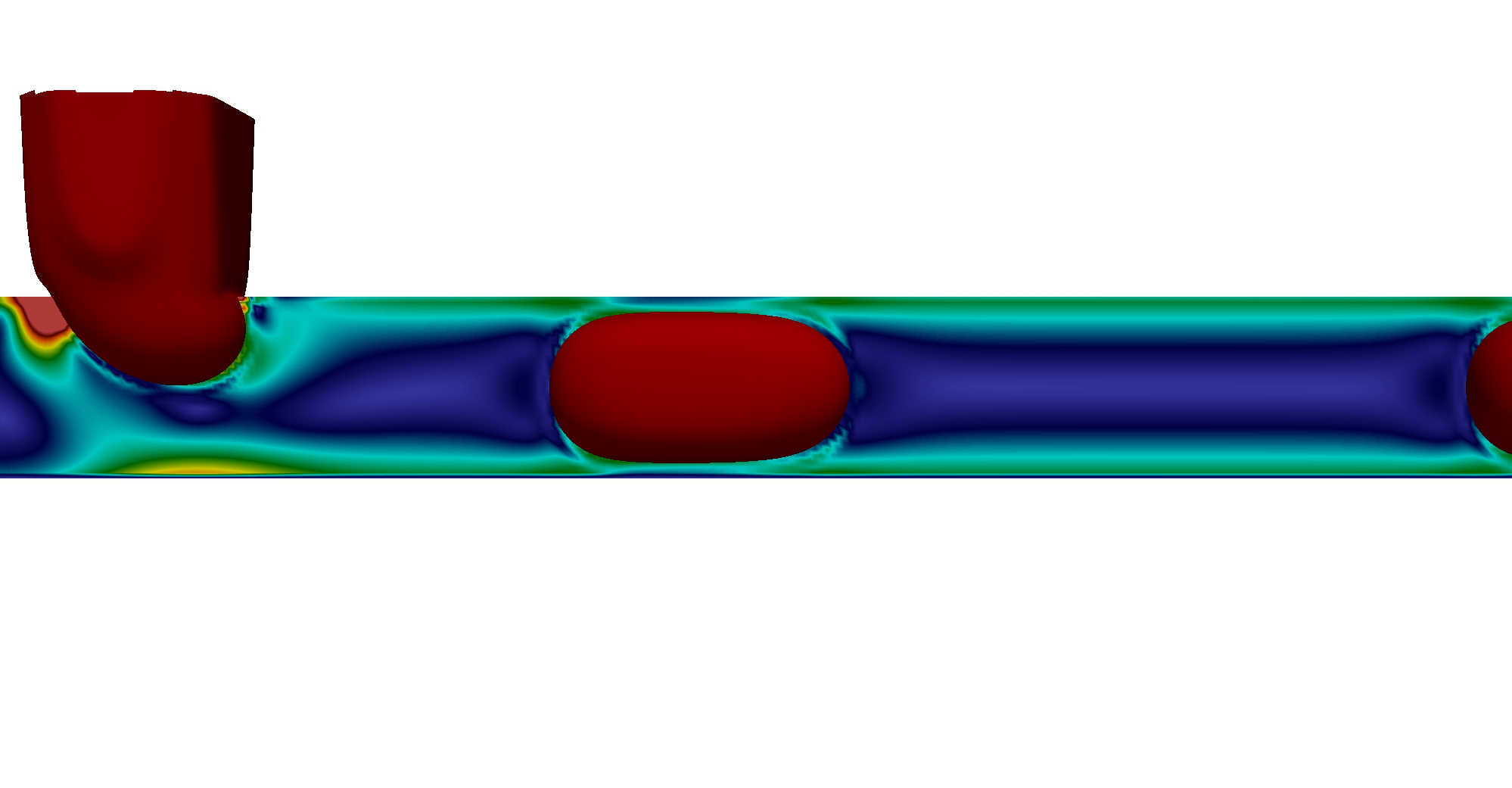}}
\caption{Viscous stress evaluated from the velocity profiles shown in Fig.~\ref{fig:Qc_choreo_PROFILE}. Further information is provided in the movies of the SM.\label{fig:Qc_choreo}}
\end{center}
\end{figure*}
%
%
%
%
\section{Conclusions}\label{sec:conclusions}
We have extensively studied droplet breakup in a microfluidic T-junction driven by either Newtonian or non-Newtonian (shear thinning) continuous phases.
The droplet size is measured over a wide range of the viscosity ratio $\lambda$ and flow rate ratio $\varphi$ still partly unexplored~\cite{Demenech07} \cite{LiuZhang11,christopher_experimental_2008}. 
Squeezing, dripping, and jetting regimes are identified for Newtonian and non-Newtonian continuous phases and the resulting breakup maps look quite similar. The droplet length in the squeezing and dripping regimes is found to nicely scale with an effective Capillary number, which reduces to the usual Capillary number when the fluid is Newtonian.
At sufficiently high $Q_c$, close to the dripping to jetting transition, where the breakup is dominated by the shear stress, droplets generated in Xanthan solutions are more elongated with respect to the Newtonian ones.
The experiments are complemented with numerical simulations based on lattice Boltzmann models (LBM) with purely thinning fluids in the squeezing-to-dripping transition. The simulations help in clarifying on a more quantitative basis the observed decrease in droplet size for the same injection flow rate. In such conditions, simulations indeed show that the viscous stress is more intense close to the microchannel walls for a shear thinning fluid, thus yielding smaller droplets, as typically observed in the experiments. 
The rescaling with the effective Capillary number is also verified in numerical simulations, thus confirming that the observed properties are solely ascribed to a combination of continuum hydrodynamics and purely thinning phases. 
Our results provide new insights into the formation and the manipulation of droplets in the presence of non-Newtonian confined environments and show that LBM can be successfully employed for the simulation of such complex microfluidic systems.
Measurements are currently in progress on the generation of oil droplets in polyacrylamide (PAA) solutions~\cite{rafai2005spreading,varagnolo17_softmatter}, characterized by strong elastic effects, while showing only weak shear thinning properties, a situation complementary to this study. Another interesting aspect for future studies could be the assessment of the validity of the effective Capillary number to rescale and predict droplet size upon changing the channel geometry, while keeping the same flow rate magnitudes.
%
%
%
%
\\
\\
{\bf Acknowledgments.}
The research leading to these results has received funding from the European Research Council under the European Community's Seventh Framework Programme (FP7/2007-2013) / ERC Grant Agreement N. 279004 (DROEMU). We acknowledge the computing hours from ISCRA B project (COMPDROP), CINECA Italy. We are grateful to Ladislav Derzsi and Davide Ferraro for fruitful discussions, and to Fabio Bonaccorso for support in the numerical simulations. We thank Stefano Bertoldo and Stefano della Fera for help in preliminary measurements.
E.C. and A.G. are equally contributed to this work.
%
%
%
%
%
%
%
\bibliographystyle{apsrev4-1-custom}
\bibliography{sample_prf-NO-URL}
\end{document}